# Converting ECG and other paper legated biomedical maps into digital

A.R. Gomes e Silva, H.M. de Oliveira, R.D. Lins

Federal University of Pernambuco - UFPE, Signal Processing Group, C.P. 7.800, 50.711-970, Recife - PE, Brazil. (hmo@ufpe.br, rdl@ufpe.br)

**Abstract.** *This paper presents a digital signal processing tool developed using Matlab$^{TM}$, which provides a very low-cost and effective strategy for analog-to-digital conversion of legated paper biomedical maps without requiring dedicated hardware. This software-based approach is particularly helpful for digitalizing biomedical signals acquired from analogical devices equipped with a plottingter. Albeit signals used in biomedical diagnosis are the primary concern, this imaging processing tool is suitable to modernize facilities in a non-expensive way. Legated paper ECG and EEG charts can be fast and efficiently digitalized in order to be added in existing up-to-date medical data banks, improving the follow-up of patients.*

## 1  Background and Set-up

Digital equipments are nowadays largely preferred to analogical ones especially due to their high-quality and flexibility of working with their output. Medical equipments that use digital technology have emerged as a true revolution in signal acquisition, analysis and diagnosis [1]. Today, electrocardiograms, electroencephalograms, electromyogram and other biomedical signals are all digital. Digital signals allow very high signal processing capabilities, easy storage, transmission and retrieval of information. The well-recognized advantages of digital technology turns it the first- choice. One of the limiting factors of adopting the digital technology is the high cost of some modern digital equipment, overall some medical ones. This is a serious barrier to be crossed by those who already have a working analogical device and/or face budget limitations. An alternative to device replacement is adopting an A/D-converter and a suitable interface to a digital microcomputer or laptop. This would also allow digitizing legated analogical data, something of paramount importance in many areas, overall in medicine as the history of patients would be kept and case studies may be correlated, etc. A number of laboratories, medical institutes and hospitals have only available analogical equipments, particularly those equipped with plottingters. The storage of these signals is rather inefficient and the data processing unfeasible. In this challenging scenario, a substantial advance can be performed by designing acquisition cards with interface to microcomputers, instead of purchasing sophisticated high-cost computerized equipments. This kind of up-grade can be beneficial to small laboratories with modest resources. Nevertheless, it is not a trivial task to assemble or to design a set-box to convert signals. This study describes the development of a software tool intended to convert a version of signals and/or spectra digitalized by a scanner (files of the extension .jpg .tif .bmp etc.) to a data file, which can be efficiently processed and stored. It deals with an alternative approach to the classical A/D conversion without requiring any specific hardware.

## 2  An A/D Image-to-Data Converter Algorithm

Many relevant but old data are only available in a chart-format and the appending new data may be suitable. For instance, this is precisely what happens in many long standing time series. How to perform efficiently such a procedure? The following description is strongly based on ECG, but it can easily be adapted to other signals, either biological or not. An implementation of the A/D platform on MATLAB$^{TM}$ is presented [2], exhibiting a few cases to illustrate the lines of the procedure.

 S1. Digitalization of the paper strip
 S2. Image binarization

    S3. Skew correction
    S4. Salt-and-pepper filtering
    S5. Axis identification
    S6. Pixel-to-vector conversion
    S7. Removing the header and trailer of the acquired signal (used for device tuning)
    S8. Splitting the ECG chart and re-assembling it.

**S1. Digitalization of the Paper Strip by Scanner up to 600 dpi**

The digitalizing process of the paper containing the chart to acquire the data can be carried out at different resolutions. Higher resolutions turn feasible details identification on the acquired image, but most applications require only a resolution high enough to achieve acceptable digitalization quality, claiming as small storage and scanning time as requested. Tests were performed over ECG charts scanned at 100, 200, and 300 dpi. The paper strip scale is in millimetres, thus a 100 dpi resolution would theoretically be sufficient to record the signal information. However, in a number of cases data from the ECG map was not retrieved at low resolution (100 dpi) mainly due to the existing similarity between the axis and the plotting trace. Besides that paper folding scanned at 100 dpi may give rise to signal discontinuities.

**S2. Image Binarization**

Image binarization is a process to translate a colour image into a binary image. It is a widespread process in image processing, especially for images that contain neither artistic nor iconographic value. Since binarization reduces the number of colours to a binary level, there are apparent gains in terms of storage, besides simplifying the image analysis as compared to the true colour image processing [3]. In this analysis, we have first discarded the axis composing the image and then applied the binarization process by Otsu's algorithm [4], since it has been shown to provide satisfactory results in many applications [3],[5]. Binarization is an important step in moving from a biomedical map image towards a digital signal, as the target of the process described herein is to obtain a sequence of values that correspond to the amplitude of a uniform time series (see S6, below).

**S3. Skew Correction**

A distortion frequently found in scanning processes is the skew caused by the position of the paper on the scanner flatbed. This rotation makes hard the analysis of data embedded in the image and increases the complexity of any sort of automatic image recognition. Whenever extracting data from a digitalized chart even 0.5 degrees or less can introduce errors on the extracted data. The algorithm presented in [6] was used here to correct the skew of the image taking as reference the axis or the border of the paper strip.

**S4. Salt-and-Pepper Filtering**

Salt and pepper noise is characterized by the presence of isolated white and black pixels in a black-and-white image [7] [8]. It may bring technical hitches in analysis of the data. In order to avoid a false identification of those pixels as piece of the analyzed chart, a filter was implemented to extract isolated pixels in the middle of a 3×3 matrix and a 3×2 matrix.

**S5. Axis Identification**

Several different types of data plotting were analysed for the sake of the generality of the methodology proposed herein. The piece of paper used to register the data may contain a grid, a box, one horizontal line and one vertical line, or no axis, but in each case a specific analysis needs to be performed in order to adequately interpret the value of the signal obtained, compensating offsets, etc.

**S6. Pixel-to-Vector Conversion**

For a better data retrieving from the image, vectors have twice the number of columns of pixels analyzed, because some peaks are found as successive vertical black lines for low resolution images. Each component of the vector is a complex number. The two components related to a same column have identical real part, which is the column index of the pixel matrix. The two consecutive imaginary parts (same real part) quantify the upper and lower limits of a vertical black line. For instance, the 2D-vector V=[…; 11+25i; 11+26i; …] has coordinates meaning that the vertical black line spans from line 25 to line 26 at the column 11. Figure 1 deal with an ECG chart with no axis. As an example, a stretch of the vector used to plotting Figure 1(b) is V = [10+26i; 10+26i; 11+25i; 11+26i; 12+25i; 12+26i; 13+26i; 13+27i 14+26i 14+28i 15+28i…].

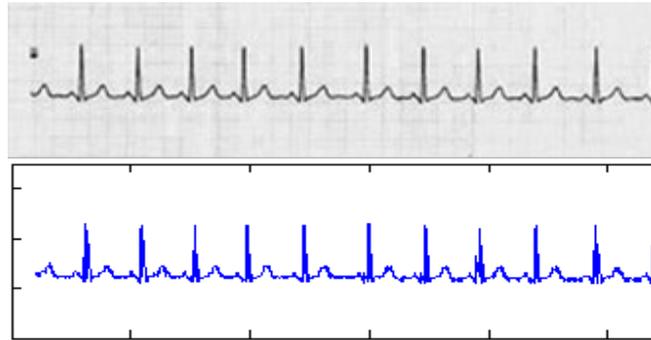

**Fig. 1.** An ECG with no axis: (a) original scanned ECG, (b) MatlabTM plotting of the ECG from the retrieved data.

The algorithm searches the data since the first pixel (bottom-left) until the last one (top-right). When there are no axis present in the map (no clear vertical and horizontal bounds), the value of the bottom vertical and the first column positions are used as a reference for ECG scanning. If step 5 (S5) is able to find axes of the graph, then these axes are assumed as data scanning reference. In order to convert the 2D-vector into one dimension, the algorithm computes the modal distance (α) between the imaginary part of two consecutive components, which means the amount of vertical black pixels composing the signal at a specific column. If the difference between the imaginary components of the two coordinates is within the limit α, then only the imaginary part of the second element is stored in the 1D version of the vector. This value is assumed as a reference (β) to the peak identification in next column analysis. If this difference is greater than α, the algorithm calculates the module of the difference between the β and each of the two components. The stored 1D-value is the one that gives the greater value. Whenever leading with plotters, one often finds portions of the chart where the drawing is no longer continuous. In order to provide a one-dimension vector, the components at those places are estimated through linear interpolation. Data retrieving can be performed from a broad range of medical-related plottings. Figure 2 shows an example of a nuclear magnetic resonance (MNR) spectrum. Horizontal and vertical lines differ from the box only by the fact that there is no upper bound in data acquisition. In appendix to this paper one may find additional examples of data extracted from ECG paper strips and the corresponding plotted data signal.

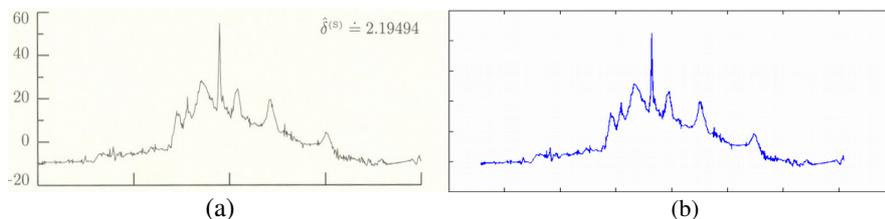

(a)                                                               (b)

**Fig. 2**. Example of data from a nuclear magnetic resonance (MNR) spectrum (extracted from the public domain software package Wavelab). (a) Original scanned spectrum, (b) MatlabTM plotting from the corresponding data file.

**S7. Removing the Header and Trailer of the Acquired Signal**

Typical ECG and a large number of ordinary graphs are composed either by boxes or horizontal and vertical lines the same colour as the plotting. Whenever those lines encompass parts of the area of the graphic, it becomes rather difficult to remove the interference, since it can happen above and/or below the graphic.

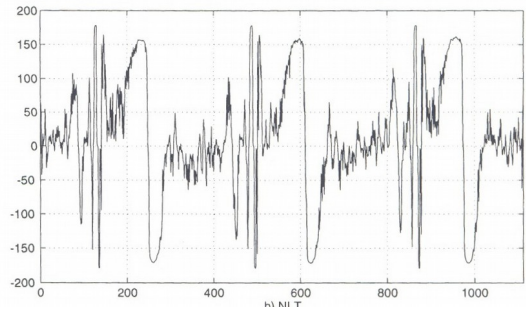

**Fig. 3**. Image of an ECG in a box, having some horizontal and vertical dotted lines.

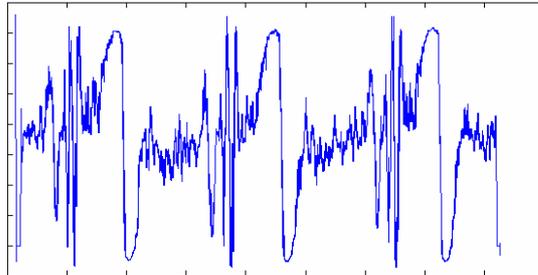

**Fig. 4**. Plotting of signal vector corresponding to the image from Fig. 3 with header and trailer removed.

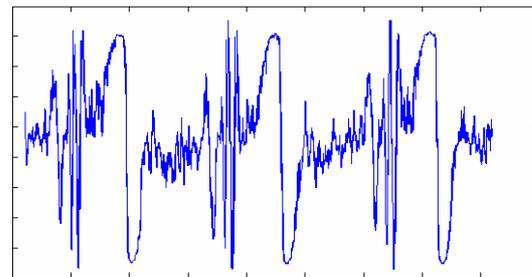

**Fig. 5**. Plotting of data retrieved from Fig. 3 after throwing away the first and final 16-values of the acquired data vector.

As the process of data acquisition is done by searching vertically, only the vertical lines interfere with the process. Besides that, very often at the beginning and ending of the data acquisition process a line is drawn carrying no information. Such headers and trailers carry no real signal information, thus they are removed from the generated vector signal, as illustrated in the example presented in Figs. 3 to 5.

**S8. Splitting the ECG Chart and Re-assembling it**

A problem that arises whenever scanning an ECG and other legated paper data signals is that usually the scanner flatbed does not cover the whole length of the paper strip, leading the operator to scan it in separated parts. Special care and image processing is needed to avoid loss or redundancy in data. A possible solution to this problem is to insert in the paper strip easily detectable marks. During the tests, handmade marks were inserted on the paper strip, drawn by pen with no other tool or mechanical support. The mark should be of a colour not originally present in the paper data strip. The image processing algorithm scans the image horizontally for one or two of such marks: only one mark on the first and on the last stretch, or two marks on the intermediary ones. Those marks play the role of the horizontal bounds in step 6. The vector is extracted from the beginning of the image until the mark for the first stretch, from the first mark to the second one for the intermediary stretches and from the mark until the end of the image for the last stretch. After obtaining the vectors of each partial ECG, one can handle data by appending segments.

## An Application of Signal Processing Tools to ECG Paper Chart

This section illustrates the application of the wavelet decomposition [9] to the ECG shown in Fig. 2A (Appendix). The aim here is to corroborate that it is feasible to handle a data file derived from a strip of paper through this digitalization approach. No comments about the signal analysis are made: we just check whether or not usual signal processing techniques can be applied to such kind of retrieved data without abnormal or unexpected results. The algorithm proposed in this paper was first used to extract a Matlab data file as to allow the wavelet toolbox to be used straight away [2]. The acquired image was converted into a 1D vector that was loaded using the wavemenu command. As an example, a three level decomposition of this ECG using the Haar wavelet is shown in Figure 6.

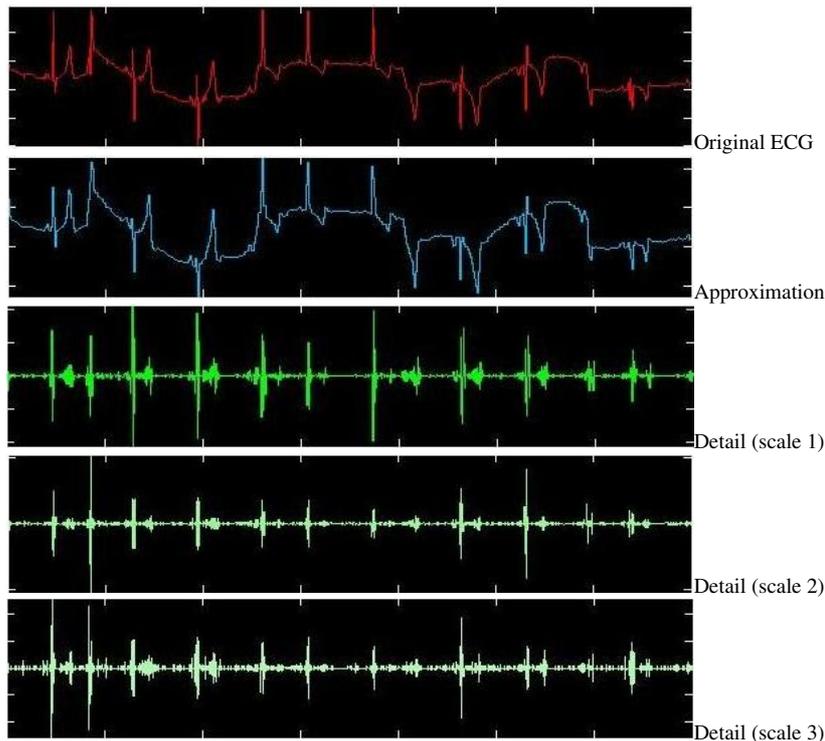

**Fig. 6**. Example of an ECG processing using the Haar wavelet corresponding to ECG paper strip of the Fig. 2A. Analysis was performed from the data acquired via the proposed algorithm.

## CLOSING REMARKS

Notwithstanding the fairly amount of scientific results, this paper describes the foundation of an efficient tool to generate digital data signals from legated paper charts. The solution proposed is a low-cost software tool that can be particularly helpful to scientists and engineers. In particular, research institutes, laboratories, clinical centres, hospitals and medical offices can largely have benefit of this up-and-coming technique, particularly due to its user-friendliness, cost-effectiveness, and accuracy. One still can save data as Matlab$^{TM}$ file or as ASCII files and edit or complement the data. One natural follow-up step to the tool presented herein is to generalise the processing capability to multi-plotting paper charts, that also frequently appears in legated data. Besides that at a later stage we hope to generalise the tool to work with signals recorded paper disks.

**ACKNOWLEDGEMENTS**. The authors are grateful to Mr. Bruno T. Ávila for making available some C command lines.

## Appendix

In this appendix we present a couple of the data extracting from true ECG strips [10] and the plottings of the retrieved data. (Fig. 1A,2A): we first find out the grid by using the blue colour information, and then switch the pixel by white pixel.

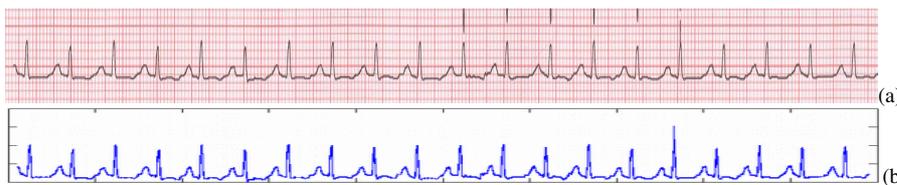

**Fig. 1A.** Image of an ECG with red grid: (a) original ECG chart, (b) Matlab[TM] plotting of retrieved ECG with data file available.

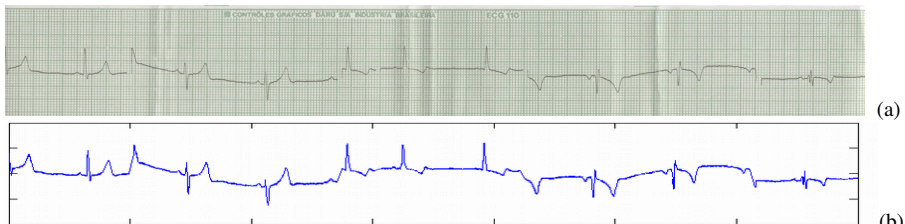

**Fig. 2A.** A green grid ECG: (a) original ECG chart, (b) Matlab[TM] plotting of retrieved ECG with available data file.

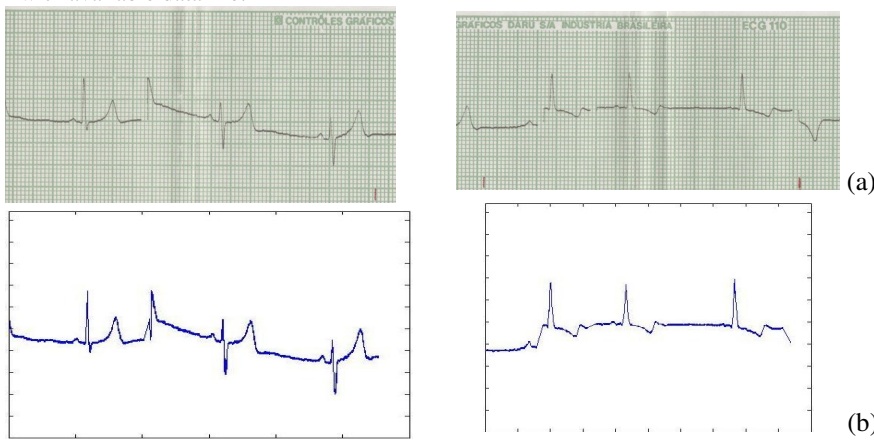

**Fig. 3A.** Image of an ECG: (a) Two ECG charts with red handmade marks drawn by pen (initial and intermediary portion), (b) Matlab[TM] plotting of retrieved ECG with data file available.